\documentclass[10pt,preprint]{aastex}

\newcommand{\HEF}{HE\,1327$-$2326}
\newcommand{\HEC}{HE\,0107$-$5240}

\shorttitle{3D LTE abundance analysis of extreme halo stars}
\shortauthors{Collet et al.}

\begin{document}

\title{The chemical compositions of the extreme halo stars 
	{\HEC} and {\HEF} inferred from 3D hydrodynamical model atmospheres}

\author{R. Collet}
\affil{Department of Astronomy and Space Physics, Uppsala University, BOX 515, SE--751 20 Uppsala, Sweden}
\email{remo@astro.uu.se}
\author{M. Asplund and R. Trampedach} %\altaffilmark{1}
\affil{Research School of Astronomy \& Astrophysics, 
The Australian National University, Mount Stromlo Observatory, Cotter Road, Weston ACT 2611, Australia}

\begin{abstract}
We investigate the impact of realistic 3D hydrodynamical model stellar
atmospheres on the determination of elemental abundances in 
the carbon-rich, hyper iron-poor stars {\HEC} and {\HEF}.
We derive the chemical compositions of the two stars by means 
of a detailed 3D analysis of spectral lines under the assumption of
local thermodynamic equilibrium (LTE).
The lower temperatures
of the line-forming regions of the hydrodynamical models cause
changes in the predicted spectral line strengths.
In particular we find the 3D abundances of C, N, and O 
to be lower by $\sim -0.8$~dex (or more) than estimated
from a 1D analysis. 
The 3D abundances of iron-peak elements are also decreased but
by smaller factors ($\sim -0.2$~dex). We caution however that the neglected
non-LTE effects might actually be substantial for these metals. 
We finally discuss possible implications
for studies of early Galactic chemical evolution. 

\end{abstract}


\section{Introduction}
\citet{christlieb02} and \citet{frebel05} recently
reported the discovery of two extreme halo stars,
the red giant {\HEC} and the subgiant or dwarf {\HEF},
with an iron abundance more than 100\,000 times lower 
than the solar one. 
While the abundances of iron-peak elements in {\HEC} and {\HEF}
are the lowest ever observed in stellar objects, these 
stars are also remarkable as that they show 
very large overabundances of carbon, nitrogen, and oxygen 
with respect to iron.
The lively interest aroused by {\HEC} and {\HEF} comes 
from the consideration that they could be directly related to the very
first generation of metal-free (Population III) stars 
to form in the early Universe.

Various mechanisms have been proposed to explain the
origin and chemical composition of these two stars.
In particular the question has been raised whether {\HEC} 
and {\HEF} could actually be Population III stars or 
whether instead they were born from the ashes of a 
previous generation of metal-free stars.
Possible scenarios of the first kind include self-enrichment 
\citep{picardi04,weiss04}, accretion from the interstellar
medium \citep{shigeyama03}, or mass transfer 
from a close asymptotic giant branch (AGB) companion
star \citep{suda04}.
However, these mechanisms have been shown to be unlikely or
only partially satisfactory in reproducing the composition 
of both {\HEC} and {\HEF} \citep[see discussions in][]{christlieb04, aoki06}.

Alternatively, the two stars might have formed 
out of material polluted by just one single or a few 
first-generation stars.  
\citet{limongi03} suggested that the 
chemical composition of {\HEC} might be naturally 
explained by the combined yields from a low mass 
(${\sim}15$~M$_{\sun}$) and a massive (${\sim}35$~M$_{\sun}$) 
supernova.
\citet{umeda03} proposed instead a scenario with 
a single ${\sim}$25~M$_{\sun}$ Population III star 
undergoing supernova explosion and experiencing mixing after 
the explosive nucleosynthesis and subsequent fallback 
on the compact remnant. 
\citet{iwamoto05} argued that small variations
in the explosion energy of this supernova model 
could in fact reproduce 			
the chemical composition of both {\HEC} and {\HEF}.
Finally, \citet{meynet06} recently showed that 	
winds of rotating massive primordial stars 
might account for the C, N, and O excesses
and possibly also for the Na and Al enhancements
in these as well as in other extremely metal-poor stars. 

Accurate determination  of the chemical compositions of
{\HEC} and {\HEF} is crucial for the identification of
the most plausible formation scenarios for these stars.
Ordinary assumptions and approximations in classical 
stellar abundance analyses involve 1D, 
local thermodynamic equilibrium (LTE),
hydrostatic model atmospheres 
relying on a simplified treatment of convective energy 
transport. 
However, recent 3D time-dependent simulations of stellar surface 
convection indicate that the structural 
differences between 3D hydrodynamical and 1D hydrostatic
model stellar atmospheres have a significant impact 
on the predicted strength of spectral lines and hence on 
the derived abundances for stars with very low metallicity 
\citep{asplund99,asplund01}.
Here we present the results of an 
abundance analysis of {\HEC} and {\HEF} based on 3D hydrodynamical
model atmospheres.

\section{Methods}
\subsection{Convection simulation}
We use the 3D, time-dependent,
radiative-hydrodynamical code by \citet{stein98} 
to carry out the first ever surface convection simulation of a metal-poor red giant star.
The stellar parameters correspond to \HEC: T$_{\rm eff}\simeq5130$~K, $\log{g}=2.2$~[cgs],
and a scaled solar composition \citep{grevesse98} with $\rm{[X/H]}$\footnote{
[A/B]$\equiv\log(n_{\mathrm{A}}/n_{\mathrm{B}})-\log(n_{\mathrm{A}}/n_{\mathrm{B}})_\sun$,
where $n_{\mathrm{A}}$ and $n_{\mathrm{B}}$ are the number densities of
elements A and B respectively and subscript $\sun$ refers to the Sun.}$=-3$
for all metals. 
As shown by \citet{christlieb04}, the thermal structure 
of a 1D model atmosphere with $\rm{[Fe/H]}=-3$ very closely 
resembles the one from a 1D model atmosphere tailored to the
specific chemical composition of {\HEC}.
This essentially reflects the dominating role
of hydrogen both as opacity source as well as
electron donor in metal-poor stellar atmospheres;
also, the too high iron-peak abundances
are partly compensated by the too low CNO abundances. 
We therefore expect our differential analysis 
to still provide a satisfactory estimate of 3D$-$1D effects
on spectral line formation in {\HEC}.

The hydrodynamical equations of mass, momentum, and energy
conservation are solved together with the 3D radiative transfer 
equation on a Eulerian mesh with 
100$\times$100$\times$125 grid-points. 
The physical domain of the simulation is large enough 
($\sim1150\times1150\times450$~Mm$^3$) to cover about ten granules 
simultaneously and twelve pressure scales 
in the vertical direction. In terms of continuum optical depth 
at $\lambda=5000$~{\AA} the simulation extends from 
$\log{\tau_{\rm{5000}}}\la-5$ down to $\log{\tau_{\rm{5000}}}\ga7$.
The temporal evolution of the simulation covers several convective
turn-over time-scales to allow for thermal relaxation.
For the simulations we employ open boundaries vertically and
periodic boundaries horizontally.
At each time-step  the 3D radiative transfer
equation is solved along one vertical and eight inclined rays.
The opacities are grouped in four opacity bins \citep{nordlund82} 
and LTE is assumed throughout the calculations.
The adopted equation-of-state comes from \citet{mihalas88} and
accounts for the effects of ionization, excitation and 
dissociation of 15 of the most abundant elements, as well as
the H$_2$ and H$_2^+$ molecules.
Continuous opacities come from the Uppsala opacity package 
\citep[and subsequent updates]{gustafsson75} and line opacity
data from \citet{kurucz92,kurucz93}.

\begin{figure}
\plotone{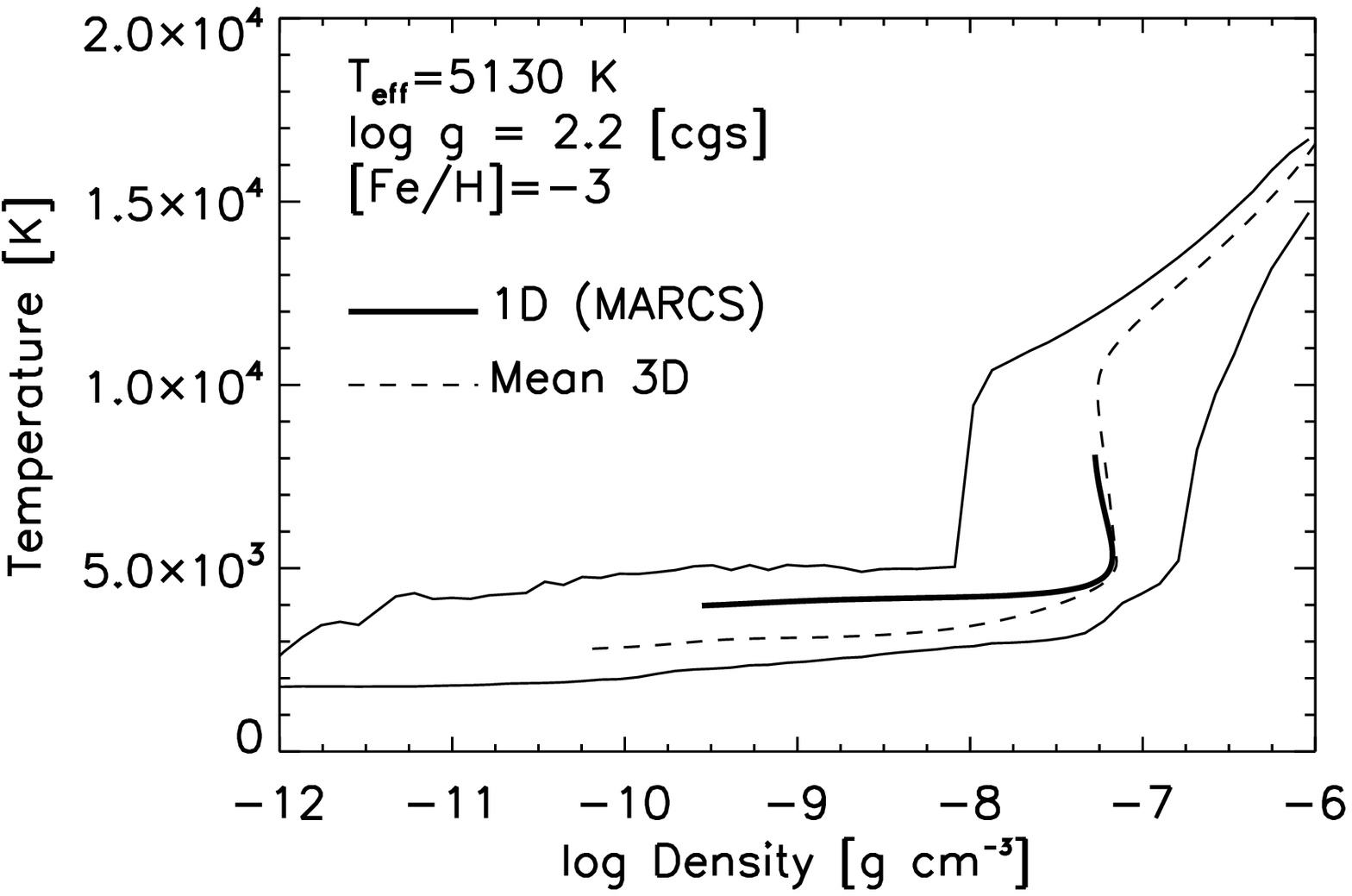}
\caption{ 
Thin solid curve: extreme temperatures at a given density 
in the 3D model atmosphere of metal-poor giant.
Thin dashed line: mean temperature-density stratification for 
the 3D model. 
Thick solid line: temperature-density stratification 
for the corresponding {\sc marcs} model.\label{atmos1}}
\end{figure}

The thermal structure resulting from the convection simulation 
is shown in Fig.~\ref{atmos1} together with the temperature 
stratification from a classical 1D, plane-parallel, hydrostatic
{\sc marcs} model atmosphere \citep{gustafsson75,asplund97} constructed 
for the same stellar parameters and chemical composition
(and a micro-turbulence of $\xi=2.0$\,km\,s$^{-1}$). 
Similarly to what was found by \citet{asplund99} and \citet{asplund01}
for metal-poor dwarfs and subgiants, in the hydrodynamical 
simulation the temperature of the surface layers tends to remain much 
lower than in the 1D model atmosphere where radiative equilibrium 
is enforced.
The temperature in the optically thin layers of the simulation 
is for the most part regulated by two competing mechanisms: 
adiabatic cooling following the expansion of the ascending gas 
and radiative heating by spectral lines. 
With fewer and weaker lines available at low metallicities,
adiabatic cooling becomes more dominant and the balance between 
cooling and heating is reached at lower surface temperatures
\citep{asplund99}.

\subsection{Spectral line formation} \label{specline}
We use the red giant convection simulation as a time-dependent 
3D hydrodynamical model atmosphere to study detailed spectral 
line formation under the assumption of LTE.
While both our 3D hydrodynamical simulation and 1D {\sc marcs} model 
atmosphere are constructed for a metallicity  $\rm{[X/H]}=-3$,
in the line formation calculations we assume the 
chemical composition to be the same as for {\HEC} 
\citep{christlieb04,bessell04} when computing ionization
and molecular equilibria and continuous opacities.
This is necessary since adopting a 
composition with metallicity $\rm{[X/H]}=-3$ would 
overestimate the abundance of elements with low
ionization potentials (e.g. Na, Ca, Al) and therefore
the electron density, affecting ionization balance and, ultimately, line strengths.

We also evaluate differential 3D$-$1D effects 
in the hyper iron-poor star {\HEF} \citep{frebel05} by
adopting a previous 3D model \citep{asplund01} of 
metal-poor turn-off star 
(T$_{\rm eff}\simeq6200$~K, $\log{g}=4.04$~[cgs], and $\rm{[Fe/H]}=-3$)
and the corresponding 1D {\sc marcs} model atmosphere. 
While this $\log{g}$ is intermediate between the current best 
estimates for {\HEF}, $\log{g}=3.7$ or $\log{g}=4.5$, the derived
abundances are only marginally
sensitive to the choice of surface gravity in that range.
Furthermore, the 3D$-$1D abundance corrections we compute are 
even less dependent on the exact value of $\log{g}$.
For the line formation
calculations we assume the chemical composition of
{\HEF} derived by \citep{aoki06}.
For the C, N, and O abundances we adopt values
midway between the subgiant and dwarf solution.

From the full red giant and turn-off star simulations 
we select two representative sequences, respectively 
8\,000 and 60 minutes long, of about 30 snapshots separated 
at regular intervals in time.
Prior to the line formation calculations 
we decrease the horizontal resolution  
from $100\times100$ down to $50\times50$ and increase the 
vertical resolution of the layers with $\log\tau_{\rm{5000}}\la3$ 
to improve the numerical accuracy.  

We compute spectral line profiles for all lines 
from neutral and singly-ionized metals considered 
in the 1D analyses of \citet{christlieb04} and 
\citet{aoki06} (see also Table~\ref{tabions}).
For each line we solve the radiative transfer equation 
along 33 directions (4 $\mu$-angles, 8 $\phi$-angles, and the vertical),
after which we perform a disk-integration and a time-average over 
all selected snapshots.
To estimate the impact of 3D models, we derive 
elemental abundances (or their upper limits) from 
the measured equivalent widths and carry out a  
differential comparison with the corresponding 1D {\sc marcs} model
atmospheres. 
We also consider a set of weak ``fictitious''
spectral lines representative of features in
CH, C$_{\rm{2}}$, CN, OH, and NH molecular bands 
\citep{christlieb04,bessell04,aoki06,frebel06} 
with typical lower-level excitation potentials 
and $\log{gf}$ values (Table~\ref{tabmol}). 
We determine the 3D$-$1D corrections to C, N, and O abundances
by carrying out a similar comparison of equivalent widths 
for these lines with 3D and 1D models.
While for the 1D analysis we adopt a micro-turbulence 
$\xi=2.2$\,km\,s$^{-1}$ for the red giant and 
$\xi=1.6$\,km\,s$^{-1}$ for the turn-off star,
we emphasize that no micro- nor macro-turbulence 
parameters enter the 3D spectral line synthesis calculations:
the velocity fields inherent to the hydrodynamical 
simulations are sufficient to 
reproduce the line broadening associated with 
convective Doppler shifts \citep{asplund05}.

\section{Results}
The 3D$-$1D LTE abundance corrections for atomic lines are presented 
in Table~\ref{tabions} and Fig.~\ref{xfe}.
Lines of different species possess 
varying sensitivities to the temperature structure, 
therefore magnitudes and signs of the corrections 
depend on the considered transition.
As 3D model metal-poor stellar atmospheres 
are significantly cooler at the surface than their 1D 
counterparts, the fraction of neutral-to-ionized metals 
is typically enhanced in the 3D simulations.
Consequently, for a given abundance, low-excitation lines 
of neutral minority species (e.g.~\ion{Fe}{1}) 
appear stronger in the framework of 3D 
models than they do in 1D, resulting in negative 
3D$-$1D corrections.
Low-excitation lines of majority species also (e.g.~the
\ion{Ca}{2}\,H\,\&\,K lines, \ion{Sr}{2}, \ion{Ba}{2}, and 
\ion{Eu}{2} transitions considered here) 
have negative 3D$-$1D abundance corrections: 
the effect in this case is not due to increased line opacities 
but rather to the decrease in continuous opacity 
associated with the lower density of $\rm{H}^-$ ions in 
3D model atmospheres of metal-poor stars \citep{asplund05}.
High-excitation lines on the contrary, like the \ion{S}{1}, 
\ion{Fe}{2}, and \ion{Zn}{1} features examined here, 
form in deeper photospheric layers and are essentially 
insensitive to the lower surface temperatures encountered 
in the 3D models.  The resultant 3D$-$1D abundance corrections for these 
lines are relatively small and mostly positive.

In Table~\ref{tabmol} we present the relative 3D$-$1D
abundances of C, N, and O.
Molecule formation in late-type stars is highly sensitive 
to the temperature of the upper photospheric layers.
The density of molecules is greatly
enhanced in 3D models with respect to the 1D case, 
leading to large negative 3D$-$1D LTE abundance corrections.
Similarly as for neutral metals,
these corrections are more pronounced the lower the
excitation potential of the molecular line.

Overall we find for {\HEF} very large 3D$-$1D corrections 
to the C, N, and O abundances that can reach $-1.0$~dex 
for the lowest-excitation CH, NH, and OH lines.
Concerning {\HEC}, \citet{christlieb04}
found a discrepancy of $0.3$~dex between the 1D LTE carbon abundance
values determined from CH and C$_{\rm{2}}$ molecular lines.
Our 3D analysis brings the C abundances derived from 
these two indicators down to a consistent level 
of $\log{\epsilon(\rm{C})}{\simeq}5.7$~dex.
Assuming the above value for the C abundance,
we find substantial 3D$-$1D corrections to the 
N abundance derived from CN lines ($\sim\,-1.7$~dex or even larger).
Based on the recent identification of NH molecular bands 
in the UV spectrum of {\HEC} \citep{bessell05}
we compute 3D$-$1D corrections to the N abundance
derived from these lines and find them to be significantly
smaller than for the CN features ($\sim\,-1.0$~dex).
The 3D N abundance derived for {\HEC} from CN
lines is ${\sim}0.8$~dex lower than the one inferred from
NH lines (Table~\ref{tabmol}).
While the two indicators also yield discrepant N abundances
in the 1D analysis, the agreement %between the 3D estimates
is significantly worse in 3D.
However, the origin of this discrepancy is likely ascribable to 
inaccurate determination of the physical parameters of NH lines. 
\citet{bessell05} and \citet{aoki06} actually \emph{decreased}
Kurucz's $\log{gf}$ values of these lines by $\sim-0.5$ dex
upon calibration with the solar spectrum; 
\citet{spite05} on the other hand observed
in their analysis of extremely metal-poor giants
that NH lines with Kurucz's $\log{gf}$ values
give systematically ($\sim+0.4$~dex) \emph{higher} N abundances 
than CN lines.

\section{Discussion}
\begin{figure}
\plotone{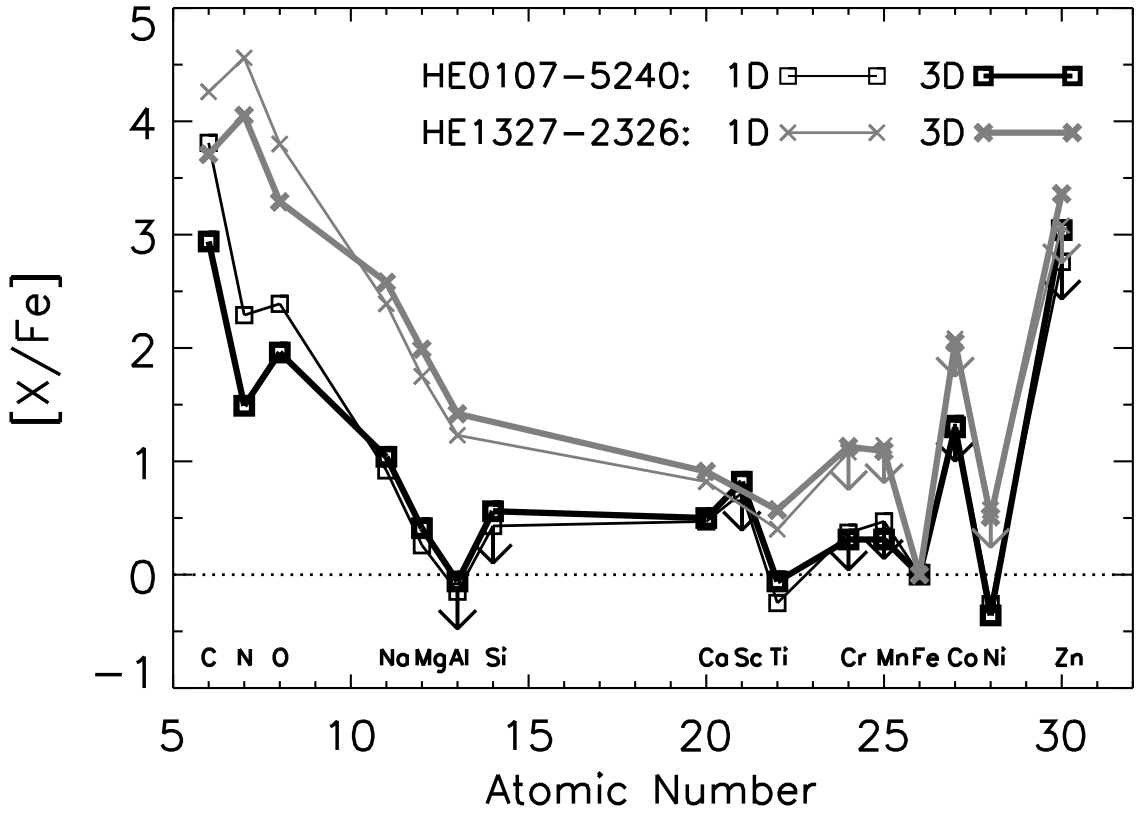}
\caption{ Comparison of 1D LTE (thin lines) and 3D LTE (thick lines) 
elemental abundance ratios in {\HEC} (squares) and {\HEF} (crosses);
arrows indicate upper limits. 
The N abundances in {\HEC} and {\HEF} here shown are
both inferred from NH lines.
The new solar abundances by \citet{ags05} are used. 
The adopted 3D$-$1D LTE corrections for the CNO abundances are averages
over low-excitation molecular lines. 
\label{xfe} }
\end{figure}

The impact of 3D hydrodynamical models on the
derived elemental abundances of extreme halo stars 
is of significance for our  understanding 
of the early phases of Galactic chemical evolution.
The most remarkable 3D$-$1D LTE effect on the abundance ratios
(Fig.~\ref{xfe}) is a severe reduction of the 
C, N, and O enhancements in {\HEC} and {\HEF}.
This result suggests that the CNO yields of first generation
stars might actually be systematically lower than previously thought.
In particular the revised 3D abundances of carbon appear  
in closer agreement with the [C/Fe] ratio predicted by
\citet{karlsson06} with a stochastic Galactic chemical evolution model
even when relatively low C stellar yields \citep{meynet02} are adopted.
Also, as rotating massive stars are believed to be the main contributors of
primary nitrogen in the early Galaxy, 
our downward revision of N abundance in hyper iron-poor stars
can have strong implications for the modelling of the
structure and evolution of primordial stars.

Overall, the lower CNO enhancements resulting from the
3D analysis could be a challenge for the
formation scenario proposed by \citet{iwamoto05}
given that the 3D$-$1D effects on Na, Mg, and Al are far
less pronounced \citep[see discussions in][]{frebel06}.
It is important to emphasize however that many of the
lines considered in the analysis of {\HEC} and {\HEF} 
are expected to suffer from non-LTE effects \citep{christlieb04,frebel05},
given the steep temperature gradient in the
atmospheric stratification and the weak UV line-blocking 
in these metal-poor stars.
A complete 3D non-LTE analysis of the two stars is
beyond the scope of the present work; however, 
we can to first approximation estimate non-LTE effects 
on \ion{Fe}{1} lines by means of a 1D analysis both
with {\sc marcs} models and mean atmospheric 
stratifications inferred from the 3D simulations. 
Using the model Fe atom by \citet{collet05} we find 
substantial non-LTE corrections due to severe 
\ion{Fe}{1} over-ionization feeding on the strong
UV radiation field, even when fully efficient 
Drawin-like \citep{drawin68,drawin69}
inelastic H$+$Fe collisions are taken into account: 
$\rm{[Fe/H]_{non-LTE}}\simeq-4.7$~dex for {\HEC} and
$\rm{[Fe/H]_{non-LTE}}\simeq-5.1$~dex for {\HEF}
These values are significantly larger than the ones
reported by \citet{christlieb04} and \citet{frebel05},
because of differences in the adopted efficiency of the H$+$Fe collisions.
The problem of non-LTE \ion{Fe}{1} line formation
in hyper iron-poor stars certainly requires further 
investigation. 
We defer the study of non-LTE effects on Fe and other
elements in 3D models to a future paper.

\acknowledgments
The authors acknowledge support from
the Swedish Foundation for International Cooperation in Research 
and Higher Education and the Australian Research Council.
K. Eriksson, B. Gustafsson, and T. Karlsson 
are thanked for valuable suggestions and fruitful discussions.

%%%%%%%%%%%%%%%%

\clearpage

\clearpage

\begin{deluxetable}{lc c rr c rr}
%\tabletypesize{\scriptsize}
\tablecaption{3D$-$1D LTE corrections applied to
the 1D abundances derived for {\HEC} and {\HEF} from metal lines. 
Abundances are expressed in the customary logarithmic scale where $\log\epsilon(\mathrm{H})\equiv{12}$.
Dots in column 2 indicate that average corrections 
over two or more lines of the same ion are reported. \label{tabions} }
\tablewidth{0pt}
\tablehead{
\colhead{} & \colhead{} & \colhead{} & \multicolumn{2}{c}{{\HEC}} & \colhead{} & \multicolumn{2}{c}{{\HEF}} \\
	\cline{4-5} 	\cline{7-8}	
\colhead{Ion} 	& \colhead{$\lambda$~(\AA)}   	& \colhead{} 	& \colhead{$\log\epsilon_{\rm{1D}}$\tablenotemark{a}} & \colhead{$\log\epsilon_{\rm{3D}}$} 
						& \colhead{} 	& \colhead{$\log\epsilon_{\rm{1D}}$\tablenotemark{b}} & \colhead{$\log\epsilon_{\rm{3D}}$}
}
\startdata
\ion{Li}{1}  & $6707.8$  &&  $< 1.12$  &   $< 0.94$   &&  $< 1.5 $  &  $< 1.42$  \\
\ion{Na}{1}  & \ldots    &&  $  1.86$  &   $  1.75$   &&  $  3.06$  &  $  3.01$  \\
\ion{Mg}{1}  & \ldots    &&  $  2.41$  &   $  2.33$   &&  $  3.63$  &  $  3.63$  \\
\ion{Al}{1}  & $3961.5$  &&  $< 0.93$  &   $< 0.79$   &&  $  2.04$  &  $  1.99$  \\
\ion{Si}{1}  & $3905.5$  &&  $< 2.55$  &   $< 2.45$   &&  $  	 $  &  $      $  \\
\ion{S }{1}  & $4034.0$  &&  $< 7.11$  &   $< 7.15$   &&  $  	 $  &  $      $  \\
\ion{Ca}{1}  & $4226.7$  &&  $  0.99$  &   $  0.84$   &&  $  0.95$  &  $  0.85$  \\
\ion{Ca}{2}  & $3933.7$  &&  $  1.44$  &   $  1.24$   &&  $  1.52$  &  $  1.37$  \\
\ion{Sc}{2}  & $3613.8$  &&  $<-1.50$  &   $<-1.63$   &&  $  	 $  &  $      $  \\
\ion{Ti}{2}  & \ldots    &&  $ -0.62$  &   $ -0.66$   &&  $ -0.24$  &  $ -0.31$  \\
\ion{Cr}{1}  & $4254.3$  &&  $< 0.65$  &   $< 0.36$   &&  $< 1.09$  &  $< 0.90$  \\
\ion{Mn}{1}  & $4033.1$  &&  $< 0.47$  &   $< 0.08$   &&  $< 0.87$  &  $< 0.58$  \\
\ion{Fe}{1}  & \ldots    &&  $  2.06$  &   $  1.83$   &&  $  1.79$  &  $  1.55$  \\
\ion{Fe}{2}  & $5018.4$  &&  $< 3.00$  &   $< 3.06$   &&  $< 3.01$  &  $< 3.07$  \\
\ion{Co}{1}  & $3453.5$  &&  $< 0.86$  &   $< 0.60$   &&  $< 1.34$  &  $< 1.07$  \\
\ion{Ni}{1}  & $3414.8$  &&  $  0.58$  &   $  0.27$   &&  $< 1.16$  &  $< 0.86$  \\
\ion{Ni}{1}  & \ldots    &&  $  0.60$  &   $  0.27$   &&  $	    $  &  $	 $  \\
\ion{Zn}{1}  & $4810.5$  &&  $< 1.97$  &   $< 2.02$   &&  $< 2.02$  &  $< 2.06$ \\
\ion{Sr}{2}  & $4077.7$  &&  $<-2.83$  &   $<-3.00$   &&  $<-1.77$  &  $<-1.88$ \\
\ion{Ba}{2}  & $4934.1$  &&  $<-2.33$  &   $<-2.59$   &&  $<-2.06$  &  $<-2.25$ \\
\ion{Eu}{2}  & $4129.7$  &&  $<-1.99$  &   $<-2.52$   &&  $  	 $  &  $      $ \\
\enddata

\tablenotetext{a}{From \citet{christlieb04}.}
\tablenotetext{b}{From \citet{aoki06}, assuming $\log{g}=3.7$~[cgs].}
\end{deluxetable}

\clearpage

\begin{deluxetable}{l ccc ll c rr}
%\tabletypesize{\scriptsize}
\tablecaption{Differential 3D and 1D C, N, and O abundances for {\HEC} and {\HEF} 
from the analysis of fictitious molecular lines (see section~\ref{specline}).
The 1D abundances are adopted from the literature, %\citet{christlieb04,bessell04,bessell05,aoki06,frebel06};
while the 3D abundances are the ones that reproduce the equivalent widths
of the fictitious lines calculated with 1D models.
\label{tabmol}}

\tablewidth{0pt}
\tablehead{
\colhead{} & \colhead{} & \colhead{}  & \colhead{} & \multicolumn{2}{c}{{\HEC}} & \colhead{} & \multicolumn{2}{c}{{\HEF}} \\
	\cline{5-6} 	\cline{8-9}	
\colhead{Species} & \colhead{$\lambda$~(\AA)} & \colhead{$\chi$~(eV)}	& \colhead{} & 
	\colhead{$\log\epsilon_{\rm{1D}}$} & \colhead{$\log\epsilon_{\rm{3D}}$} & \colhead{} & \colhead{$\log\epsilon_{\rm{1D}}$} & \colhead{$\log\epsilon_{\rm{3D}}$} 
}
\startdata
C~(CH) & $4360$	& 0.0 &&  $6.81$    &  $5.72$	  &&  $ 6.77 $  &  $5.96$ \\
       &   	& 0.2 &&  $\ldots$  &  $5.79$     &&  $\ldots$  &  $6.03$ \\
       &   	& 0.3 &&  $\ldots$  &  $5.82$     &&  $\ldots$  &  $6.07$ \\
       &   	& 0.5 &&  $\ldots$  &  $5.90$     &&  $\ldots$  &  $6.13$ \\
C~(C$_2$)&$5160$& 0.0 &&  $7.11$    &  $5.63$	  &&    &  \\
         &	& 0.2 &&  $\ldots$  &  $5.72$     &&    &  \\
         &	& 0.3 &&  $\ldots$  &  $5.76$     &&    &  \\
         &	& 0.5 &&  $\ldots$  &  $5.86$     &&    &  \\
N~(CN) & $3880$ & 0.0 &&  $5.22$\tablenotemark{a} ($4.93$\tablenotemark{b})
				    &  $2.91$\tablenotemark{c}     &&    &  \\
       &	& 0.2 &&  $\ldots$  &  $3.04$\tablenotemark{c}     &&    &  \\
       &	& 0.3 &&  $\ldots$  &  $3.12$\tablenotemark{c}     &&    &  \\
       &	& 0.5 &&  $\ldots$  &  $3.27$\tablenotemark{c}     &&    &  \\
N~(NH) & $3360$	& 0.0 &&  $4.83$    &  $3.78$	  &&  $ 6.58 $  &  $5.76$ \\
       &   	& 0.2 &&  $\ldots$  &  $3.85$     &&  $\ldots$  &  $5.83$ \\
       &   	& 0.4 &&  $\ldots$  &  $3.92$     &&  $\ldots$  &  $5.90$ \\
       &   	& 0.8 &&  $\ldots$  &  $4.04$     &&  $\ldots$  &  $6.01$ \\
O~(OH) & $3150$	& 0.0 &&  $5.66$    &  $4.95$     &&  $ 6.65 $  &  $5.80$ \\
       &   	& 0.5 &&  $\ldots$  &  $5.04$     &&  $\ldots$  &  $5.95$ \\
       &   	& 0.8 &&  $\ldots$  &  $5.08$     &&  $\ldots$  &  $6.02$ \\
       &   	& 1.0 &&  $\ldots$  &  $5.12$     &&  $\ldots$  &  $6.07$ \\
       &   	& 1.5 &&  $\ldots$  &  $5.19$     &&  $\ldots$  &  $6.20$ \\
\enddata
\tablenotetext{a}{Assuming $\log\epsilon(\rm{C})_{\rm{1D}}=6.81$}
\tablenotetext{b}{Assuming $\log\epsilon(\rm{C})_{\rm{1D}}=7.11$}
\tablenotetext{c}{Assuming $\log\epsilon(\rm{C})_{\rm{3D}}=5.71$}

\end{deluxetable}

\end{document}